\documentclass[pra,twocolumn,nofootinbib]{revtex4}
\usepackage{amssymb,amsmath}
\usepackage{graphics}
\usepackage{rotating}
\begin{document}

\title{Coulomb deexcitation of pionic hydrogen within close-coupling method}

\author{V.\,N.\,Pomerantsev, V.\,P.\,Popov}

\affiliation{Institute of Nuclear Physics, Moscow State University}

\begin{abstract}{The Coulomb deexcitation of $(\pi p)$ atom
in collisions with the hydrogen atom has been studied in the
fully quantum mechanical close-coupling approach for the first
time. The total Coulomb deexcitation cross-sections of
$nl\rightarrow n'l'$ transitions and $l$-averaged cross-sections
are calculated for $n=3 - 8$ and at the relative energies
$E=0.01 - 100$ eV. The strong interaction and vacuum
polarization shifts of $ns$-states are taken into account.
 It is shown that the $\Delta n > 1$
transitions are very important and make up a substantial
fraction of the Coulomb deexcitation cross-section (up to $\sim
47\%$).}
\end{abstract}


\maketitle

Exotic hydrogen atoms are formed in highly excited states and the
evolution of their distributions on the quantum numbers and kinetic
energy  is defined by the competition of both collisional and
radiative deexcitation processes during the atomic cascade. The
interest in the problem attracted special attention after the
experimental observation of a high energy fraction ($\gg 1$~eV) in the
energy distribution of the pionic atoms at the instant of the charge
exchange reaction $\pi^- p \to \pi^0 n$ was obtained~\cite{1,2,3,4}.
The observed Doppler broadening of the neutron time-of-flight (NTOF)
spectra in the experiments are attributed to the Coulomb deexcitation
(CD) process, where the energy of the transition is shared between the
colliding objects. In order to analyze the above mentioned and the new
experiments~\cite{5,6,7,8} on the precision spectroscopy of $K$
$X$-rays from pionic atoms, the kinetic-energy distribution of pionic
hydrogen must be calculated from a realistic cascade model.

   The first theoretical study has been carried out in the framework
of the two-state semiclassical model~\cite{9} with some additional
approximations (e.g., $n \gg 3$). Later, the process was considered
within the advanced adiabatic~\cite{10} (see also references therein),
and the classical-trajectory Monte Carlo~\cite{11} approaches. While
for the low $n$ states the approaches~\cite{9, 11} cannot be expected
to give a reliable description of the CD process, the CD cross
sections obtained in~\cite{10} for the $n\!\to \!n-1$ transitions
($n=3-5$) are smaller about one order of the magnitude than in
\cite{9} and too small to explain the experimental data~\cite{12}.
Therefore, in the most important region of $n=3-7$ relevant for the
atomic cascade the theoretical results are rather undefined and this
process is regarded as the least known process of the pionic cascade.

In this paper, we report the results for the CD process of
$\pi^- p$ atom in collisions with $H$ obtained for the
first time in the framework of the fully quantum mechanical
approach. We use the close coupling (CC) method, in which
the scattering processes
\begin{equation} \label{eq. 1}
(\pi^- p)_{nl} + H_{1s}\rightarrow(\pi^- p)_{n'l'} +H_{1s},
\end{equation}
such as elastic scattering ($n'\!=\!n,\,l'\!=\!l$), Stark transitions
($n'\!=\!n,\,l'\!\neq \!l$), and CD ($n'\!< \!n$) are described in unified manner.
The method has been recently employed for the study of the CD process
in muonic hydrogen~\cite{13}, where  substantially new results in
comparison with the known in the literature~\cite{9, 14} (see
references therein) have been obtained. Unless otherwise stated the
atomic units ($\hbar \!=\! e\! =\!m_e m_p/(m_e\!+\!m_p) \!=\! 1$) will be used
throughout the paper.

 The Hamiltonian of the system $((\pi^- p)_{nl}  + H_{1s})$
 (after separation of the c.m. motion) is given by
\begin{equation} \label{eq2}
H = -\frac{1}{2m}\Delta _{\mathbf{R}} +
h_{ex}(\boldsymbol{\rho}) + h_H(\mathbf{r})
+V(\mathbf{r},\boldsymbol{\rho},\mathbf{R}),
\end{equation}
where $m$ is the reduced mass of the system, $\mathbf{R}$ is the
radius vector between the c.m. of the colliding atoms,
$\boldsymbol{\rho}$ and $\mathbf{r}$ are their inner coordinates. The
interaction potential, $V(\mathbf{r},\boldsymbol{\rho},\mathbf{R})$,
is a sum of four Coulomb pair interactions between the projectile atom
and the target atom particles. $h_{ex}(\boldsymbol{\rho})$ and
$h_H(\mathbf{r})$ are the hydrogen-like Hamiltonians of the free
exotic and hydrogen atom, whose eigenfunctions together with the
angular wave function $Y_{L\Lambda}(\hat{\bf R})$ of the relative
motion form the basis states, $|1s, n l,L:JM\rangle$, with the
conserving total angular momentum ($J M$) and parity $\pi=(-1)^{l+L}$.
The total wave function of the system can be expanded as follows
\begin{equation} \label{eq3}
\Psi^{J M \pi}_E(\mathbf{r}, {\boldsymbol{\rho}},
\mathbf{R}) = R^{-1} \sum_{nl L}G_{nlL}^{J \pi}(R) |1s, n
l,L:J M\rangle.
\end{equation}

The expansion (3) leads to the close-coupling second order
differential equations for the radial functions of the
relative motion, $G_{nlL}^{J \pi}(R)$,
 \begin{multline}  \label{eq4}
\left(\frac{d^2}{dR^2} + k^2_{n} -
\frac{L(L+1)}{R^2}\right)G^{J \pi }_{nlL}(R) = \\
=2m\sum_{n'l'L'}W^{J \pi}_{n'l'L', nlL}(R)\, G^{J
\pi}_{n'l'L'}(R).
\end{multline}
The channel wave number is defined as
$k^{2}_{n}=2m(E_{cm}+E_{n_0 l_0}-E_{n l})$, where $E_{cm}$ and
($n_0 l_0$) are the energy of the relative motion and the exotic
atom quantum numbers in the entrance channel, respectively. The
bound energy of $\pi p$ atom, $E_{n l} = \varepsilon _{nl} +
\Delta \varepsilon _{nl},$ includes the eigenvalueof
$h_{ex}(\boldsymbol{\rho})$, $\varepsilon _{nl}$,  and the
energy shift, $ \Delta \varepsilon _{nl}=\Delta \varepsilon
^{vp} _{nl}+\Delta \varepsilon ^{str}_{nl}$, due to the vacuum
polarization and strong interaction. Hereafter, the energy
$E_{cm}$ will be referred to $\varepsilon _{n l\neq 0}$ in the
entrance channel (we assume that $\Delta \varepsilon _{n
l\neq 0}=0$)\footnote{The strong interaction shifts are
calculated according to $\Delta \varepsilon ^{str}_{ns}=\Delta
\varepsilon ^{str}_{1s}/n^3$, where $\Delta \varepsilon _{1s}=-
7.11$ eV from \cite{16}). The next values~\cite{17}  ($- 3.24$~eV
($1s$), $- 0.37$~eV ($2s$), and $- 0.11$~eV ($3s$)) for the vacuum
polarization shifts are used in the calculations. For $n>3$ we
assume $\Delta \varepsilon ^{vp} _{ns}=\Delta \varepsilon ^{vp}
_{n-1,s}((n-1)/n)^3$.}.

The matrix elements of the interaction potential,
\begin{equation} \label{eq5}
W^{J}_{n'l'L', nlL}(R)=\langle 1s,n'l',L':JM|V|1s,nl,L:JM\rangle
\end{equation}
are obtained by averaging it over the electron wave function of
the $1s$-state and then applying the multipole expansion. The
integration over (${\boldsymbol{\rho}}, \hat{\bf R})$ reduces
the matrix elements (5) to the multiple finite sum.

At fixed $E_{cm}$ the coupled differential equations (4) for the
given $J$ and $\pi$ values are solved numerically by Numerov
method with the standing-wave boundary conditions involving the
real symmetrical $K$-matrix related to $T$-matrix by the
equation $T=2K/(I - iK)$. All exotic atom states with $n$ values
from 1 up to $n_0$ have been included in the close-coupling
calculations. The following cross sections will be discussed:

- partial cross section
\begin{equation}
\sigma^J_{nl\to n'l'}(E)  =
\frac{\pi}{k_{n}^2}\frac{2J+1}{2l+1} \sum_{ L L'
\pi}|T^{J\pi}_{nlL\to n'l'L'}|^2,\,  \label{eq6}
\end{equation}
- total cross section of the transition $nl\to n'l'$
\begin{equation}
\sigma_{nl\to n'l'}(E)  = \sum_{J}\sigma ^J_{nl\to
n'l'}(E),  \label{eq7}
\end{equation}
- total cross section from $n s$ - state
\begin{equation}
\sigma_{ns\to n'}(E)  = \sum_{l'}\sigma_{ns\to n'l'}(E),
\label{eq8}
\end{equation}
and the  $l$-averaged cross section ($l>0$)
\begin{equation}
 \sigma^{l>0}_{nn'} (E) = \frac{1}{n^2-1} \sum_{l>0, l'}(2l+1)\sigma_{nl\to n'l'}(E).
   \label{eq9}
\end{equation}
We assume the pionic atom states with $l\neq 0$ are
degenerated and so the $l$-averaged cross section (9) can be used
to illustrate the basic features of the CD process.

 The CC calculations of the CD cross sections Eqs. (6-9) were carried
out for $(\pi^- p)_{nl} + H$ collisions for $n=3 - 8$ and at energy
range $E_{cm}= (0.01 - 100)$ eV. Summation over the partial waves in
Eq. (6) was done up to a value $J_{max}$ until an accuracy better than
0.1\% was reached at all energies. The analysis of the $J$ dependence
of the partial cross sections $\sigma^J_{nl\to n'l'}$ shows that  most
the CD cross sections give the partial waves with rather low $J$
values as compared with the elastic scattering and Stark transition
processes. Besides, our study shows that a significant underestimation
of the CD cross sections (up to one order of the magnitude) can be
obtained if one neglects the higher multipole terms in the expansion
of the coupling matrix (5). Such a strong effect together with the
relatively low $J$ values involved in the inelastic transitions are
related with the fact that the CD process is determined by the
comparably more short-range interaction.

\begin{figure}[h!]
\includegraphics[width=0.45\textwidth,keepaspectratio]{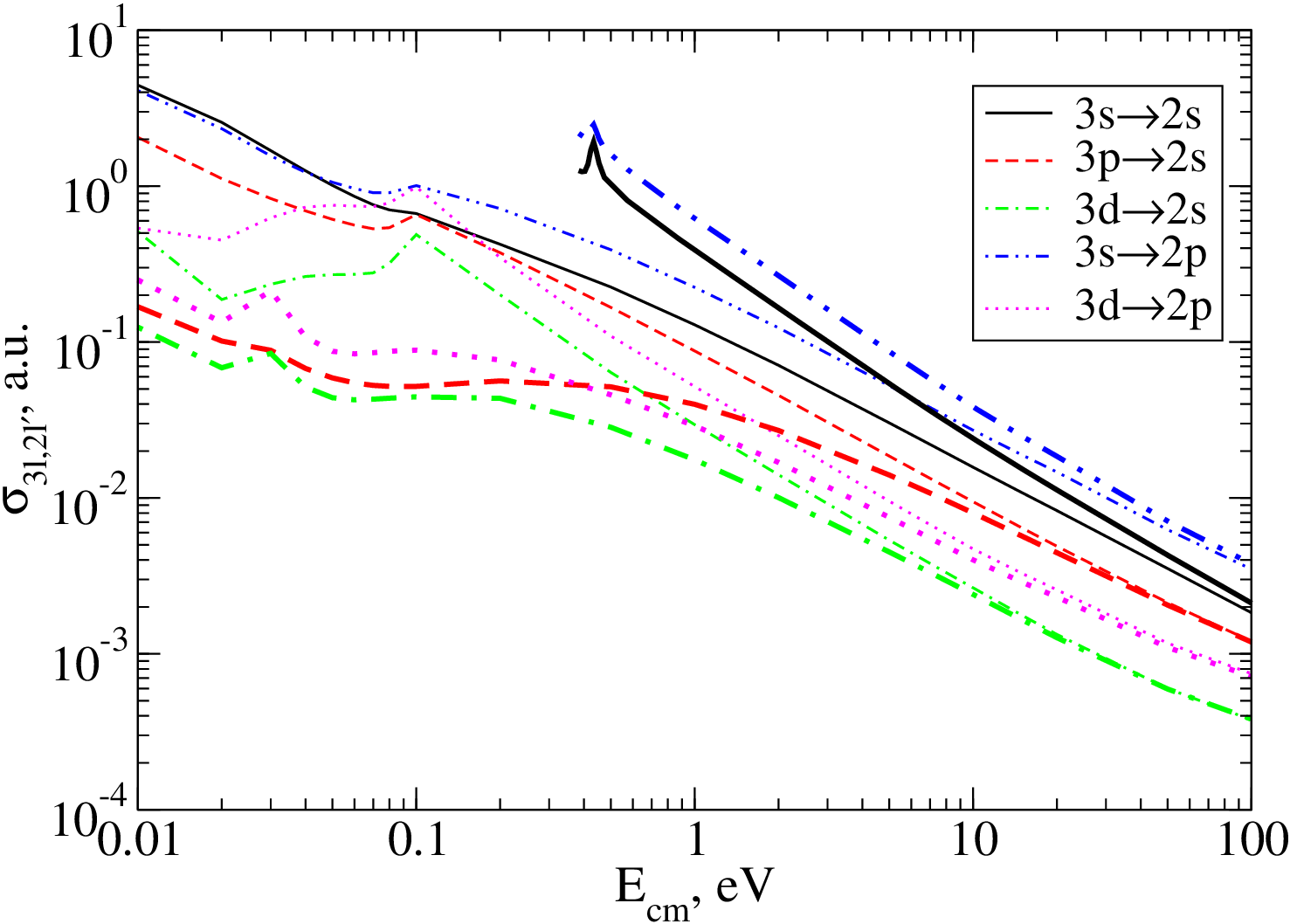}
     \caption{The CD cross sections $\sigma_{3l,2l'}$
     for $(\pi p)_{3l} + H$ collisions
 calculated both with (thick lines)  and without (thin lines) taking
 the $ns$-state energy shifts into account.}
 \label{fig1}
     \end{figure}

\begin{figure}[h!]
     \includegraphics[width=0.45\textwidth,keepaspectratio]{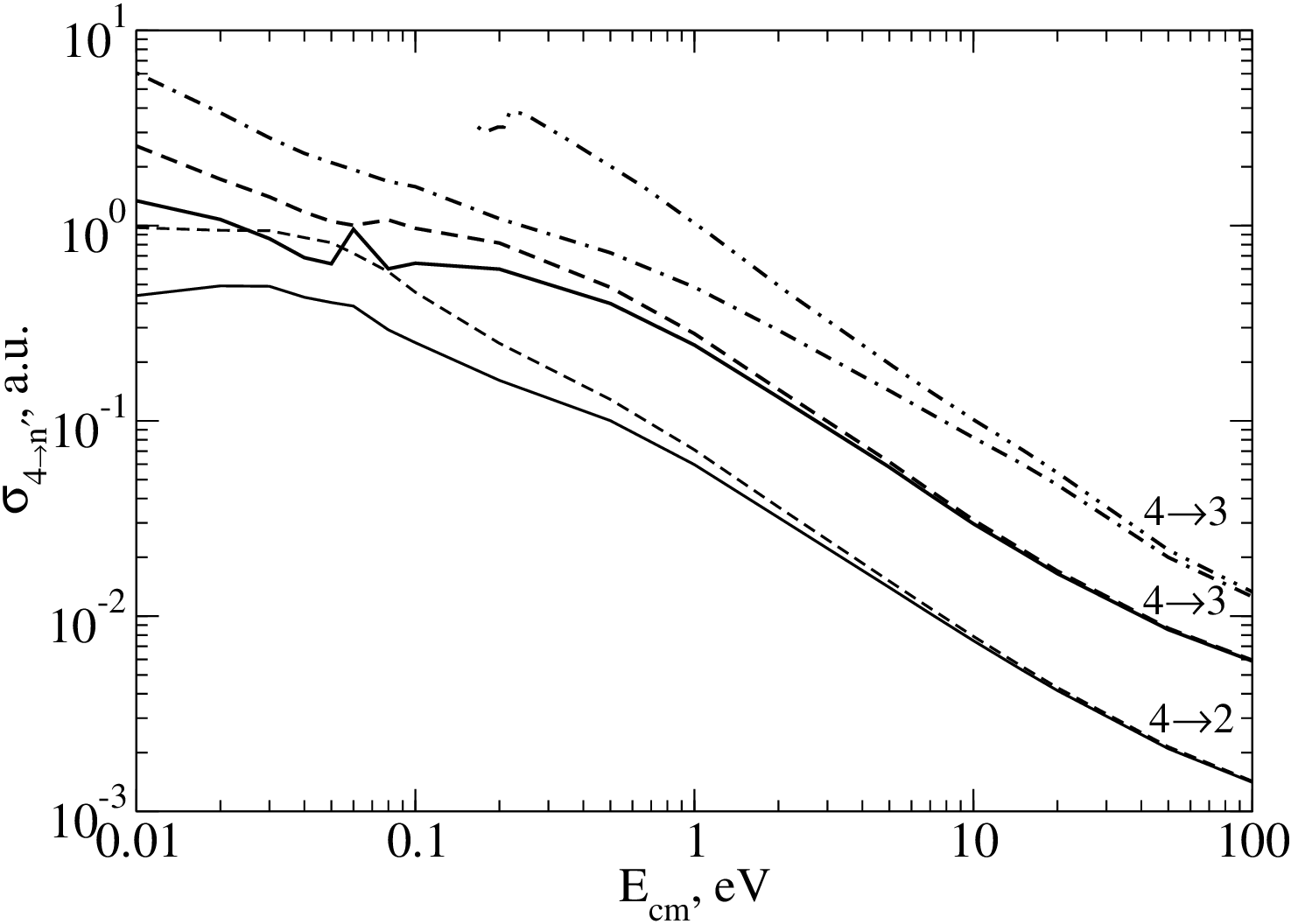}
     \caption{The $l$-averaged CD cross sections
 $\sigma^{l>0}_{4,3}$ (thick lines) and $\sigma^{l>0}_{4,2}$ (thin lines)
 for $(\pi p)_4 + H$ collisions calculated with (solid lines) and without
 (dashed lines) taking the $s$-states energy shifts into account.
 The  CD cross sections $\sigma_{4s\to 3}$ calculated both with (double-dot-dashed line)
 and without (dot-dashed line) taking
 the $s$-state energy shifts into account are also shown.}
 \label{fig2}
     \end{figure}

In order to illustrate the influence of the $ns$ state energy
shifts on the CD cross sections, the calculations were performed
both with and without taking energy shifts into account. The
effect is the most pronounced for the low-lying states and some
of the results for $n=3, 4$ are shown in Figs.~\ref{fig1},
\ref{fig2} and presented in Table~I.

The CD cross sections for a number of transitions $3 l \rightarrow 2
l'$ with or without the energy shifts of the $n s$-states are shown
in Fig.~\ref{fig1}. In the case $\Delta \varepsilon_{ns} \neq 0$ the
energy dependence of the cross sections changes crucially. The cross 
sections corresponding to the $3p (3d) \rightarrow 2s, 2p$
transitions  at energies above and below the threshold of the Stark
transitions $3s \rightarrow 3p, 3d$ are strongly suppressed. The
effect is more clearly observed at energies in the vicinity of 0.1 eV
(the suppression is more or about one order of the magnitude). At
$E_{cm} \gg |\Delta \varepsilon _{ns}|$ the differences between the
corresponding pair of the curves (see Fig.~\ref{fig1}) are negligible
at $E_{cm} \geq 15~$eV for $n=3$ and at $E_{cm}\geq 1$~eV for $n=4$.
On the contrary, the transitions $3s \to 2s, 2p$ are enhanced at the
threshold by a factor about $3 - 4$. The main  reason of such
behavior of the CD cross section  is related with the inelastic
nature of the Stark transitions between the splitted upper states.
\begin{table*}
\caption{The CD cross sections $\sigma^{l>0}_{n,n'}$
for $(\pi p)_n + H$
 collisions calculated in quantum-mechanical CC approach.}
\bigskip
\tabcolsep=0.4em
\begin{tabular}{|c|c|c|c|c|c|c|c|c|c|c|c|c|c|c|c|c|} \hline
$E_{\rm cm}$, eV& 0.01 &  0.02&  0.03& 0.05 &  0.08&  0.1 &  0.2 &  0.5 &  1.0 &  2.0 &  5.0 &  7.0 &   10 &20 &  50 & 100\\
     \hline
 $\sigma_{32}$  &0.409 &0.231 &0.289 &0.145 &0.138 &0.143 &0.139 &0.102 &0.072 &0.046 &0.022 &0.017 &0.012 &0.007 &0.003 &0.002\\
 $ \sigma_{32}^{(\Delta\varepsilon\!=\!0)}$
                &2.261 &1.315 &1.293 &1.303 &1.275 &1.711 &0.770 &0.293 &0.148 &0.075 &0.030 &0.021 &0.015 &0.008 &0.004 &0.002
\\  \hline
 $\sigma_{43}$  &1.342 &1.076 &0.858 &0.638 &0.958 &0.642 &0.599 &0.399 &0.245 &0.132 &0.058 &0.042 &0.030 &0.017 &0.009 &0.006\\
 $ \sigma_{43}^{(\Delta\varepsilon\!=\!0)}$
                &2.558 &1.728 &1.402 &1.052 &1.004 &0.969 &0.814 &0.483 &0.280 &0.145 &0.062 &0.044 &0.031 &0.017 &0.009 &0.006
\\
  $\sigma_{42}$ &0.438 &0.491 &0.489 &0.404 &0.387 &0.252 &0.161 &0.100 &0.060 &0.032 &0.014 &0.010 &0.007 &0.004 &0.002 &0.001\\
 $ \sigma_{42}^{(\Delta\varepsilon\!=\!0)}$
                &0.977 &0.946 &0.942 &0.817 &0.721 &0.456 &0.250 &0.129 &0.071 &0.036 &0.015 &0.011 &0.008 &0.004 &0.002 &0.001
\\
\hline
  $\sigma_{54}$ &3.497 &2.672 &1.750 &1.201 &1.055 &0.769 &0.531 &0.314 &0.202 &0.125 &0.060 &0.044 &0.033 &0.020 &0.013 &0.012\\
  $\sigma_{53}$ &0.929 &1.118 &0.677 &0.417 &0.355 &0.251 &0.177 &0.091 &0.050 &0.025 &0.010 &0.007 &0.005 &0.003 &0.002 &0.001\\
  $\sigma_{52}$ &0.188 &0.292 &0.199 &0.122 &0.103 &0.074 &0.069 &0.040 &0.022 &0.011 &0.005 &0.004 &0.003 &0.001 &0.001 &0.001\\
\hline
  $\sigma_{65}$ &3.644 &2.510 &2.423 &1.544 &1.121 &1.012 &0.636 &0.298 &0.169 &0.098 &0.041 &0.031 &0.024 &0.016 &0.015 &0.020\\
  $\sigma_{64}$ &1.446 &1.309 &1.232 &0.903 &0.683 &0.603 &0.377 &0.217 &0.129 &0.068 &0.028 &0.021 &0.015 &0.009 &0.005 &0.004\\
  $\sigma_{63}$ &0.436 &0.349 &0.265 &0.171 &0.115 &0.098 &0.065 &0.032 &0.017 &0.008 &0.003 &0.002 &0.002 &0.001 &0.001 &0.001\\
\hline
  $\sigma_{76}$ &5.647 &4.321 &3.202 &2.636 &2.157 &1.747 &1.247 &0.590 &0.346 &0.190 &0.082 &0.061 &0.045 &0.027 &0.023 &0.038\\
  $\sigma_{75}$ &1.545 &1.272 &0.793 &0.696 &0.762 &0.527 &0.319 &0.150 &0.095 &0.053 &0.023 &0.018 &0.014 &0.009 &0.007 &0.008\\
  $\sigma_{74}$ &0.300 &0.280 &0.154 &0.156 &0.162 &0.114 &0.069 &0.037 &0.019 &0.010 &0.004 &0.003 &0.002 &0.002 &0.001 &0.002\\
\hline
  $\sigma_{87}$ &7.480 &4.541 &4.095 &3.379 &3.217 &2.379 &1.757 &1.003 &0.662 &0.371 &0.159 &0.118 &0.086 &0.051 &0.042 &0.077\\
  $\sigma_{86}$ &1.340 &0.925 &0.822 &0.719 &0.624 &0.485 &0.378 &0.190 &0.130 &0.072 &0.031 &0.023 &0.017 &0.011 &0.010 &0.016\\
  $\sigma_{85}$ &0.571 &0.340 &0.315 &0.362 &0.234 &0.222 &0.170 &0.078 &0.044 &0.023 &0.010 &0.008 &0.006 &0.004 &0.003 &0.004\\
\hline
\end{tabular}
\end{table*}

The similar regularities are also observed in the $l$-averaged CD
cross sections. In Fig. \ref{fig2} (see also Table~1) the $l$-averaged
CD cross sections for the $4 \to 3$ and $4 \to 2$ transitions and
total cross sections from $4s$-state are presented. One can see that
the maximal suppression due to the energy shift of $4s$ state is
weaker and less or about two times at very low energy both for $4 \to
3$ and $4 \to 2$ transitions, while at $E_{cm} > 1$ eV does not exceed
15\%. The important observation follows from Fig.~\ref{fig2}:  the  CD
cross sections for the $\Delta n=2$ transition are strongly enhanced,
in contrast to the calculations~\cite{9, 10}, and their relative
contribution to the total CD transitions   $4 \to n'\leq 3$ varies
from 20\% up to 39\%.
 \begin{figure}[h!]
 \centerline{ \includegraphics[width=0.45\textwidth,keepaspectratio]
 {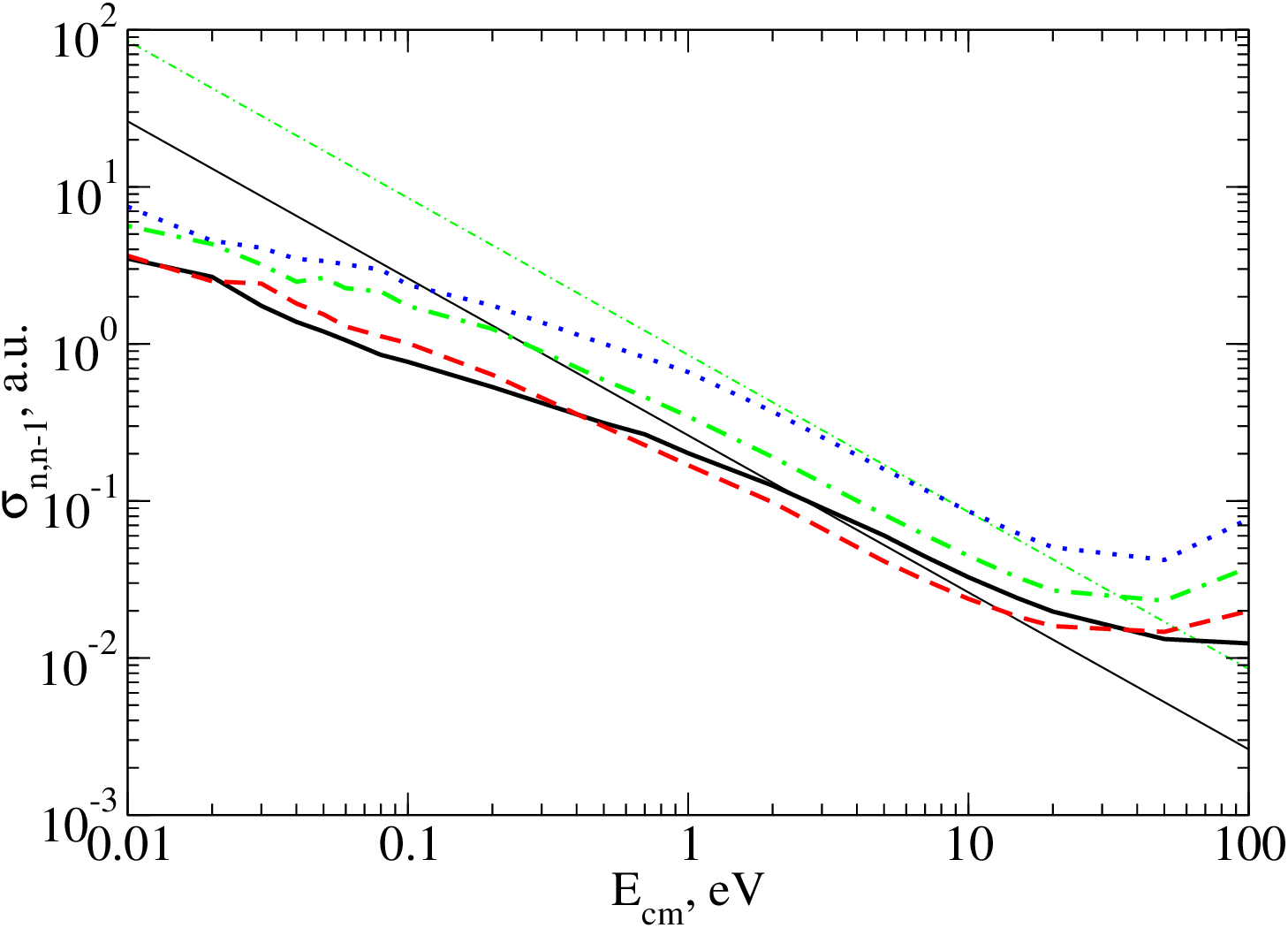} }
 \caption{The $l$-averaged CD cross sections $\sigma^{l>0}_{n,n-1}$
for $(\pi p)_n + H$ collisions with $n=5$ (solid line), 6
(dashed), 7 (dot-dashed) and 8 (dotted). The results of the
parameterization ~\cite{11} for $n=5$ and 7 (thin lines).}
 \label{fig3}
     \end{figure}

 Beginning from the paper~\cite{9}, it is commonly believed that the
CD cross sections for  $\Delta n \!= \!1$ transitions behave like ~$\sim
n^\gamma/E_{cm}$. The $n$ and $E_{cm}$ dependences of the $l$-averaged
CD cross sections ($l>0$) for $n=5 - 8$ are shown in Fig.~\ref{fig3}
for the $\Delta n \!=\!1$ transitions. The straight lines show the results
of the parameterization~\cite{11}, based on the CD cross sections
~\cite{9} for $n \leq 7$. As it is seen from Fig.~\ref{fig3}, the
energy dependence of the CD cross sections only in the region
$1\lesssim E \lesssim 10$ eV, as a whole, is in a qualitative
agreement with the one like $1/E_{cm}$. But at the energies beyond
this interval the CC results don't confirm the $1/E$ energy dependence
of the cross sections. In particular, at $E \lesssim 1$ eV and $n\!>\!4$
our results show  weaker and rather $1/\sqrt{E}$ behavior of the CD
cross sections.

Concernig the power $n$-dependence nearly to $n^{\gamma}$ with $\gamma
> 2$, the present consideration doesn't confirm that the CD cross
sections have such a scale factor depending on $n$ (see
Fig.~\ref{fig3} and Table~1). Moreover, for $n=4\div 6$ the
non-monotone behaviour of the $\sigma_n^{CD}$ as a function of $n$ is
revealed.

 \begin{figure}[h]
 \includegraphics[width=0.45\textwidth,keepaspectratio]{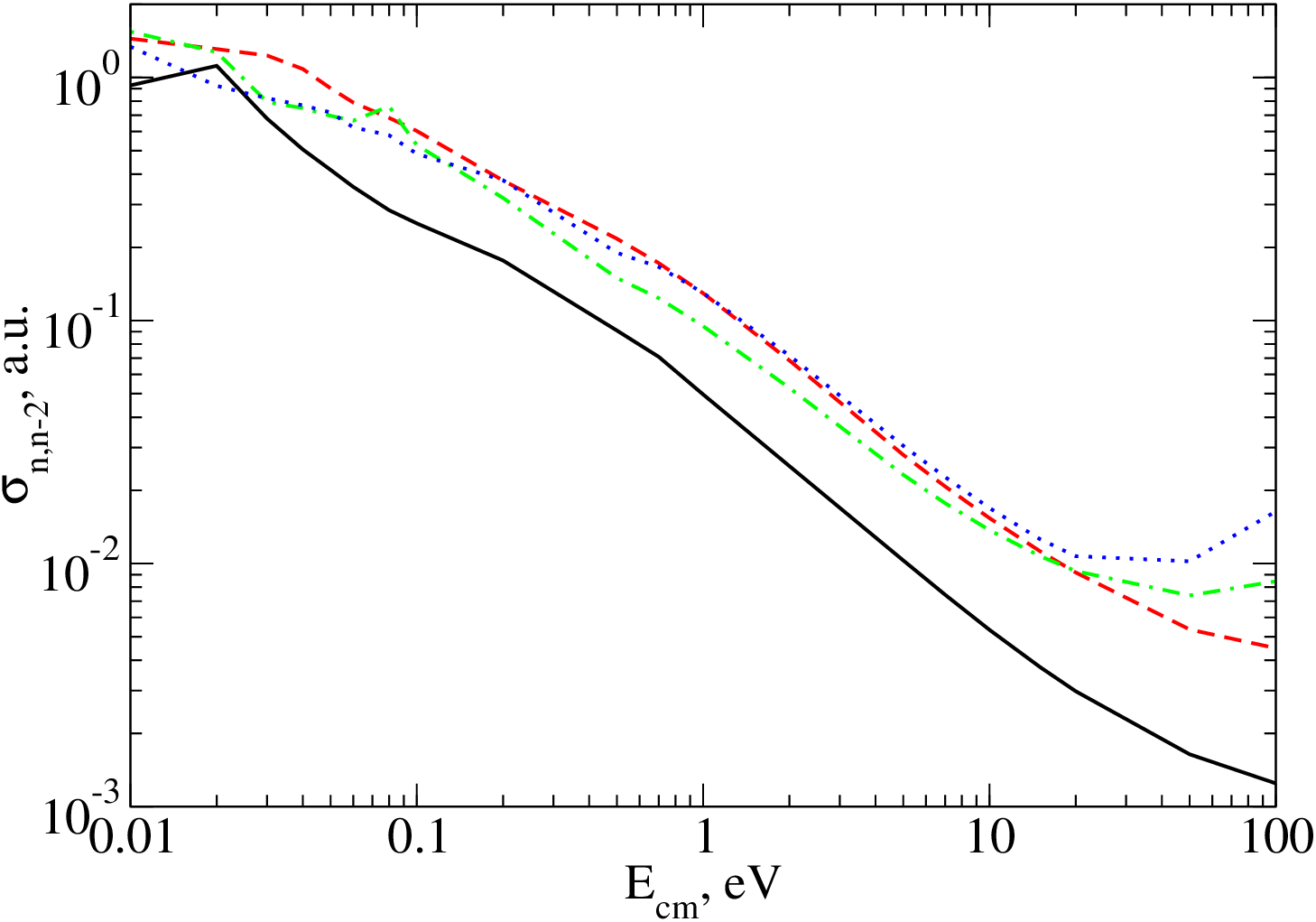}
 \caption{The same as in Fig.~3 but for transitions $n\to n\!-\!2$.}
 \label{fig4}
     \end{figure}

 \begin{figure}[h]
 \includegraphics[width=0.45\textwidth,keepaspectratio]{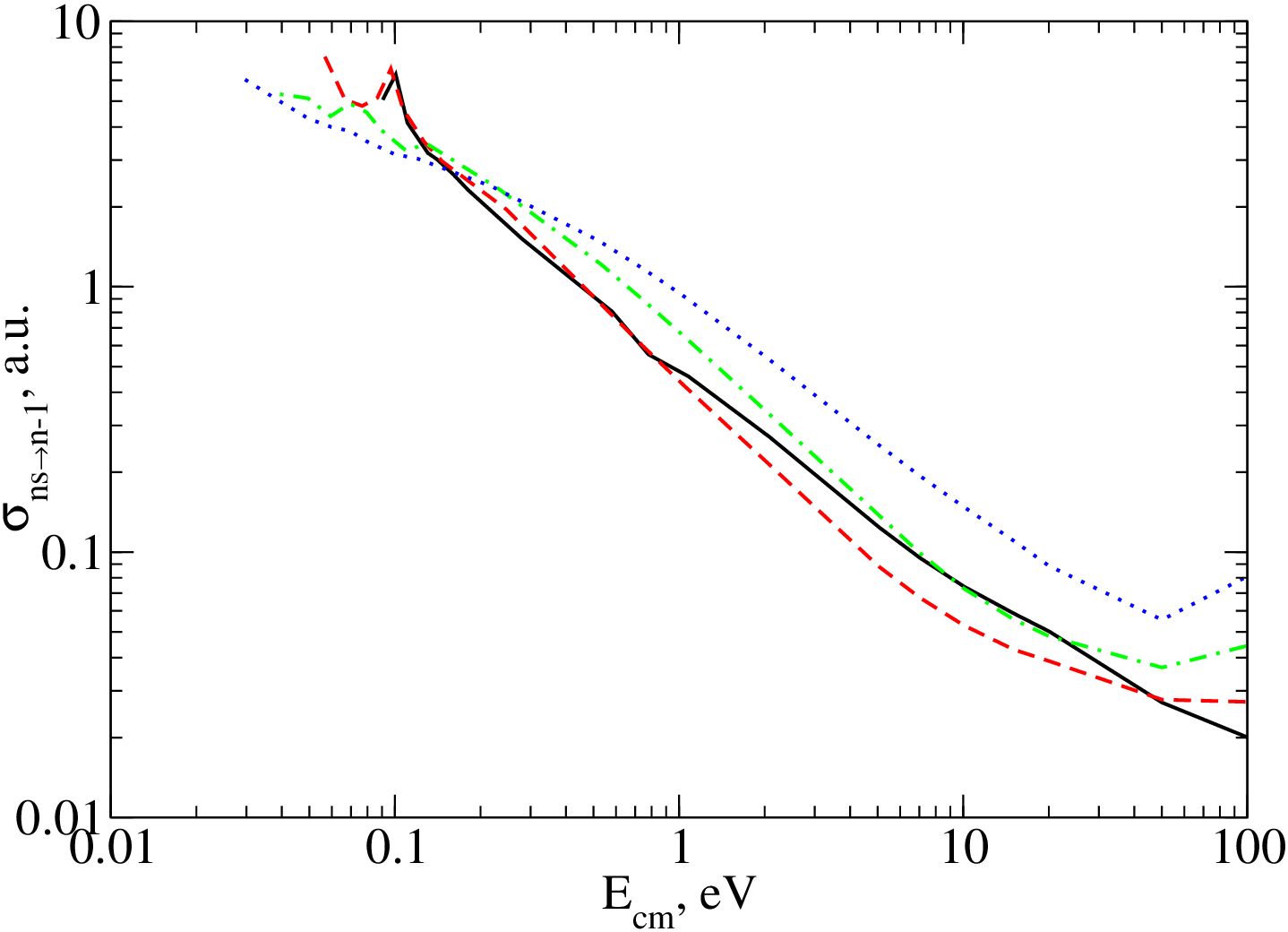}
\caption{ The CD cross sections from $s$-states $\sigma_{ns \to n-1}$
for $n=5-8$. The notations of curves are the same as in Figs.~3 and 4.}
 \label{fig5}
 \end{figure}

 The present calculations show that $\Delta n= 1$ transitions
dominate.  According to our CC calculations (see Fig.~\ref{fig4} and
Table~I), the $\Delta n>1$ transitions make up a substantial fraction
of the total CD cross sections (20--47~\%) for $n\geq 4$ at all the
energies under consideration. It is worthwhile noting, that there is
no  simple $n$ dependence of the CD cross sections for the $\Delta n>
1$ transitions. The regular $n$ dependence appears only at the
relative energies more or about 50--70~eV.

In summary, the CD process in pionic hydrogen has been studied
for the first time in the fully quantum mechanical CC approach,
involving simultaneously the elastic and Stark processes. The
obtained results reveal the new knowledge about the CD
process and are very important for the reliable analysis of the
$K$ $X$-ray yields and NTOF spectra.\footnote{We thank Dr.~ T.~
Jensen for the preliminary results of the analysis with the
present CD cross sections of the NTOF spectra~\cite{3}. Our
results explain the high energy component around 105 eV due to
the $5 \to 3$ transition and lead to a very good agreement with
the experimental weight of the $3 \to 2$ component at 209 eV.}

We are grateful to Prof. G. Korenman for fruitful discussions.
 This work was partially supported by Russian Foundation
 for Basic Research, grant No. 03-02-16616.

\end{document}